\documentstyle{mn}
\def\ut#1{\mathop{\vtop{\ialign{##\crcr     
$\hfil\displaystyle{#1}\hfil$\crcr\noalign     
{\kern1pt\nointerlineskip}\hbox{$\hfil\sim\hfil$}\crcr    
 \noalign{\kern1pt}}}}}
\def\undersim{\ut}
\def\gappreq{\undersim{>}}
\input{epsf.sty}
\begin{document}
\title{Evolution of globular cluster systems in three galaxies of 
the Fornax cluster}
\author[R. Capuzzo--Dolcetta and I. Donnarumma]
{R. Capuzzo-Dolcetta, I. Donnarumma 
\\Dipartimento di Fisica, Universit\`{a} La Sapienza, P.le Aldo Moro 2, I-00185 Roma, Italy\\
Roberto.Capuzzodolcetta@uniroma1.it, donnarum@uniroma1.it}
\date{Accepted . Received ; in original form}
\maketitle\begin{abstract}
We studied and compared the radial profiles of globular clusters and 
of the stellar bulge component in three galaxies of the Fornax 
cluster observed with the WFPC2 of the HST. A careful comparative 
analysis of these distributions confirms that stars are more 
concentrated toward the galactic centres than globular clusters, 
in agreement with what was observed in many other galaxies. 
If the observed difference is the result of evolution of the
globular cluster system sstarting from  initial profiles similar 
to those of the halo--bulge stellar components, a relevant fraction 
of their mass (74\%, 47\%, 52\% for NGC 1379, NGC 1399 and NGC 1404,
respectively) is disappeared in the inner regions, likely 
contributing to the nuclear field population, local dynamics and 
high energy phenomena in the primeval life of the galaxy.
An indication in favour of the evolutionary interpretation of the 
difference between the globular cluster system and stellar bulge 
radial profiles is given by the  positive correlation we found 
between the value of the mass lost from  the GCS and the central 
galactic black hole mass in the set of seven galaxies for which 
these data are available.
\end{abstract}\section{Introduction}
A lot of elliptical galaxies have globular cluster systems
 (hereafter GCSs) that are less concentrated to the galactic centre 
than the bulge-halo stars, thus showing radial distributions with 
larger core radii than those of the bulges. Two classical examples 
of this phenomenon are the two Virgo giants M87 and M49 
(Grillmair et al., 1986, McLaughlin, 1995), to which our Galaxy and 
M31  have been added later (Capuzzo-Dolcetta \& Vignola, 1997).
 The sample of galaxies that shows this particular feature has been 
recently enlarged thanks to 11 elliptical galaxies selected by 
Capuzzo-Dolcetta \& Tesseri (1999) in a sample of 14 galaxies with 
kinematically distinct cores (Forbes et al.,1996) observed with the 
WFPC2 of the Hubble Space Telescope. Though we could not yet 
conclude that this phenomenon is common to all galaxies, we may say 
that there is no case where the halo stars are less concentrated 
than the GCS. Moreover there is a general agreement on that the 
difference between the two radial distributions is real and not due 
to a selective bias. Consequently, different hypotheses have been 
advanced with the purpose to explain this feature. Among these, two 
seem the most probable:
\begin{enumerate}\item  the difference 
between the two distributions reflects different formation ages of 
the two systems, as suggested by Harris \& Racine (1979) and Racine 
(1991); in their opinion globular clusters originated earlier, when 
the density distribution was less peaked. However, this hypothesis 
cannot explain why the two distributions are very similar in the 
outer galactic regions.
\item  another explanation is based on the simple and reasonable
 assumption of coeval birth of globular clusters and halo stars 
(McLaughlin, 1995), with a further evolution of the GCS radial 
distribution, while the collisionless halo stands almost unchanged.
\end{enumerate}\noindent The GCS evolution is due to 
\begin{enumerate}
\item dynamical friction, that  brings massive clusters very close 
to galactic centre, 
\item tidal interaction with a compact nucleus (see for example 
Capuzzo-Dolcetta, 1993). 
\end{enumerate}
The combined effect of these dinamical mechanisms acts to deplete 
the GCSs in the central, denser regions, leaving almost unchanged 
the outer profile, that so remains similar to the halo stars' one.
The efficiency of the mentioned phenomena is higher in galactic 
triaxial potentials (see for example Ostriker et al., 1989, Pesce, 
Capuzzo-Dolcetta \& Vietri, 1992), in which we find a family of 
orbits, the `box-like', that do not conserve any component of their 
angular momentum and then well sample the central galactic regions. 
Pesce et al. (1992) showed as globular clusters moving on box orbits 
lose their orbital energy at a rate an order of magnitude larger 
than on loop orbits of comparable size and energy. 
These results showed as the previous evaluations of the dynamical 
friction efficiency, based on very simplified hypotheses on the 
globular cluster orbit distribution, led to an overstimate of the 
dynamical braking time scales. On the contrary, it has been
 ascertained that massive globular clusters in triaxial potentials 
could be braked during their motion so to reach the inner regions in 
a relatively short time(Capuzzo-Dolcetta, 1993). There they could 
have started a process of merging, giving origin to a central 
massive nucleus (eventually a black hole) or fed a preexistent one 
(Ostriker, Binney \& Saha, 1989; Capuzzo-Dolcetta, 1993).
Under the hypothesis that the initial GCS and halo-bulge radial 
distributions were the same, an accurate analysis of the 
observations allows an estimate of the number of ``missing clusters'' and therefore of the mass removed from the GCSs. 

Actually, McLaughlin (1995), Capuzzo-Dolcetta \& Vignola (1997) and
Capuzzo-Dolcetta \& Tesseri (1999), scaling the radial surface 
profiles of the halo stars of a  galaxy to that its   globular 
cluster system, estimated the number of missing globular clusters as
the integral of the difference between the two radial profiles. 
A study like this led, for example, Capuzzo-Dolcetta 
\& Vignola (1997) and Capuzzo-Dolcetta \& Tesseri (1999) to suggest 
 that the compact nuclei in our galaxy, M31 and M87 as well as in 
many other galaxies could have reasonably sucked a lot of decayed 
globular clusters in the first few Gyrs of life.
 The aim of this paper is an extension of previous (McLaughlin 1995; 
Capuzzo-Dolcetta \& Vignola 1997; Capuzzo-Dolcetta \& Tesseri, 1999)
 works to the globular cluster data of a sample of three elliptical 
galaxies in the Fornax cluster (NGC\ 1379, NGC 1399 and NGC 1404) 
whose HST data were taken by Forbes et al. 1998a,b, hereafter 
referred to F98a and F98b.
\section{\-THE DATA AND THE RESULTS}
In this Section we present and discuss our analysis of GCS radial 
distribution in the overmentioned galaxies in the Fornax cluster. 
The data come from recent HST WFPC2 observations (F98a, F98b) and, 
so, they take the advantage of the HST excellent resolution, which 
allows an accurate analysis of the crowded inner regions of galaxies
 at Virgo and Fornax cluster distance. Moreover it allows the 
subtraction of most of background galaxies. The available density
 data are, as we will see later, at a high level of completeness. 
\par\noindent We have approximated the observed density profiles 
with a modified ``core model''
\begin{equation}\Sigma (r) ={{\Sigma _{0}}\over  {\left[ 1+\left( {r}\over {r_{c}} \right) ^{2}\right] ^{\gamma}}} \label{1.1}
\end{equation}
where $\Sigma_0$, $\gamma $ and $r_c$ are free parameters ($r_c$ is 
called {\it core} radius, but, to be precise, it corresponds to the 
usual definiton of being  the distance from the centre where the 
surface distribution  halves its central value just when 
$\gamma = 1$.  We have chosen the free parameters in (1) minimizing 
the r.m.s error, i.e.
\begin{equation}\sigma =\frac{ \sum \left({ f_{i}-f_{0i} }\right) ^{2} p_{i}}{\sum f_{0i}^{2}p_{i}}
\end{equation}
where $f_{i}$ e $f_{0i}$ are, respectively, the fitted and the 
observed values of the surface density. We have defined the weights 
$p_{i}$ as the inverse  of the error bars of the surface density 
data. The modified core model approximates very well the GCS density data in the inner regions of the two galaxies NGC 1399 and NGC 1404 ($\sigma \sim10^{-4}),$ while the $\sigma $ value results a little higher ($\sim 10^{-3}$) for NGC 1379. As we will see in the following, this is justified by the somewhat peculiar behaviour of the light profile of the latter galaxy. By means of fitting formula (1) we estimated the number of ``missing'' clusters in a galaxy and the corresponding mass removed from its GCS. In fact, an approximated value of the mass fallen to the galactic centre is obtained through the two values, $N_{l}$ and $\langle m_{l}\rangle $, of the number and of the mean mass of the missing ~(\lq lost \rq)~ globular clusters. A priori, the determination of $\langle m_{l}\rangle $ needs a detailed evaluation of the effects of evolutionary processes on the original mass spectrum of the GCS. However,  the most relevant phenomena (tidal shocking and dynamical friction) act on opposite sides of the initial mass function then -- if the initial mass function is not too asymmetric -- we expect that the mean value of the globular cluster mass has not changed very much in time (see Capuzzo-Dolcetta \& Tesseri, 1997) . Hence we can assume the present mean value of the mass of globular clusters, $\langle m\rangle ,$ as a good reference value for $\langle m_{l}\rangle $. To estimate  
$\langle m\rangle$ it is thus necessary to assume a mass function that satisfactorily represents the present distributions of globular clusters in the three galaxies of our sample. On the basis of the observed similarity among the mass spectra of elliptical galaxies (Harris \& Pudritz 1994; McLaughlin 1994), we have obtained a single mass spectrum for the Fornax galaxies normalizing the mass spectrum adopted by McLaughlin (1995) for the Virgo giant M87. We have made this normalization using a mass to light ratio $\beta\equiv\left(M/L\right) _{V,\odot}$ equal to 1.5 for the GCS. The resulting mass spectrum is
\begin{equation}
N(m)\propto m^{-s}  \label{3}
\end{equation}with $s$ defined as
\begin{equation}
s =\left\{ \begin{tabular}{cc}$ s_{0}\equiv 0.5\pm 0.2$ & $m_{\min }\leq m\leq 1.2\times  10^{5}M_{\odot }$ \\ $s_{1}\equiv 1.65\pm 0.10$ & $1.2\times 10^{5}M_{\times}\leq m\leq 1.5\times
10^{6}M_{\odot }$ \\ $s_{2}\equiv 3.0\pm 0.3$ & $1.5\times 10^{6}M_{\odot }\leq m\leq m_{\max }$\end{tabular}\right.
\end{equation}
\noindent where $m_{min}$ and $m_{max}$ depend explicitly on $\beta$.Of course an error in the value of $s$ leads to an error in the evaluation of the mean mass $\langle m\rangle $. Then we think important to check the degree of reliability of our method by estimating (in the way described in the Appendix A) the mean error made using the mass spectrum (3) to evaluate the mass.
\subsection{ NGC 1379}NGC 1379 is an EO galaxy in the Fornax cluster studied by, e.g., Harris \& Hanes (1987), Kissler-Patig et al. 
(1997) and later F98b. Harris \& Hanes (1987) compared the radial profile of the GCS in NGC 1379 with the surface brightness of the galaxy itself as given by Schombert (1986). Their study was limited  to the radial range 5-35 kpc, and they did not detect any difference between the two profiles. This was confirmed by Kissler-Patig et al. (1997) that used ground-based data over the 3-10 kpc range. The recent analysis of F98b, based on data of the HST WFPC2, confirms the similarity between the two surface profiles in the outer regions but, at the same time, show as the two differ in the inner regions. Thanks to these latter data, that cover a radial range larger than previously available, we managed to make a comparative study between the halo stars' surface profile and the GCS one.
\subsubsection{Observed data and the modified core model}In F98b detailed plots of the radial profile of the GCS of NGC 1379 are presented, thanks to the high resolution of HST. The profiles shown in the paper are two: one (containing GCs with B $<25.5$) is complete and uncontaminated and the other (composed by objects with B$>26.5$ and without correction for incompleteness) is expected to be composed mostly by background galaxies. 

The radial profile of the brighter sample begins to decrease from $\sim 10$ to 80 arcsec ($\sim 1-7$ kpc); at 80 arcsec from the centre the profile is lost in the background (see F98b). 

Fig.1 shows the radial profile of clusters with B $<25.5$ with a background of 2.7 objects per arcmin$^{2}$ subtracted. The superposed dotted line represents the luminosity profile of the underlying galaxy as provided by Kissler-Patig et al. (1997),  scaled vertically to match the GCS profile in the outer regions. In this way the two profiles agree well at $\sim $35-70 arcsec (3-6 kpc). The logarithmic slope of the GCS profile is $\sim -2.4$ at $r\gappreq 35$ arcsec. Inwards of 30 arcsec it flattens out. 

F98b estimated a core radius of 23$\pm 6$ arcsec (2.0$\pm 0.5$ kpc) assuming it as the distance from the centre where the surface density halves its central value. In its turn, the galaxy light seems to decline with a constant slope from $\sim 10$ to $\sim 50$ arcsec, where it seems to change slope slightly. To be sure of completeness, we have initially analysed only the brighter sample (B$<25.5$): in a second analysis, we added the clusters of the fainter sample. Fitting the GCS density data with the law (1), we have  the best fit  with $\Sigma _{0}=220.62$ arcmin$^{-2},$ $\gamma =1$ and $r_{c}=23$ arcsec (solid line in Fig.1). The modified core model approximates very well the density data in the outer regions, while it seems to be an overestimate in the inner regions. According to F98b the central dip of the observed GCS radial distribution is not an artefact of their analysis, but it represents a real drop in the volume density of clusters near the nucleus of galaxy. In fact they estimated a 97\% completeness of the density data in the galactic region inwards of $\sim $10 arcsec. 

The observed decreasing behaviour toward the galactic centre of the GCS distribution of NGC 1379 seems to indicate that the dynamical depleting mechanisms are efficient, in this galaxy, up to significantly large radii.
\begin{figure}
\vspace{1pt}
\hspace{10pt}
\epsfxsize=240pt
\epsfbox{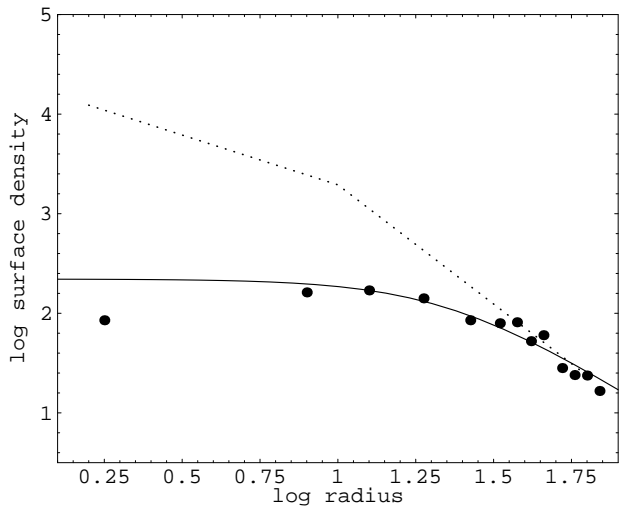}
%\epsfbox{n1379.eps}
\par\caption{Surface density profile for the the sample of globular cluster system in NGC 1379 (black dots) with $B<25.5$ from  F98b. The solid line is the modified core model fit for GCS made in our study.  The dotted curve is the surface brightness profile of the underlying galaxy, arbitrarly normalized to match the radial profile of cluster system in the outer regions.
Radial distance is in arcsec (1 arcsec = 89 pc); surface density in arcmin$^{-2}$.}
\end{figure}
\subsubsection{ The ``missing clusters'' in NGC 1379}
As we said in the Introduction, the estimate of the (potentially) missing clusters can be obtained as the integral of the difference between the GCS and halo stars' surface density distribution, after this latter has been normalized. For this reason we have  considered only the radial range where the two profiles differ, i.e. 1.77-37.58 arcsec (see Fig.1). The bulge--halo  radial profile has an almost constant slope, -2.4, in the range 10 to 50 arcsec from the centre, but for our purpose it has been necessary to know the star profile also in the region within  10  arcsec.This has been made possible thanks to Grillmair, who gave us (1999, private communication) the model used to subtract the stellar light from globular clusters profile (99 \% complete for $B<25.5$). We obtained a logarithmic slope of the stellar profile of about -1 in the radial range 1.77-10 arcsec (see Fig.1). In the annulus 1.77-37.58 arcsec, the number of stars is 512, on the other side, the modified core model fitting the globular cluster distribution gives 132 as  number of clusters observed in the same annulus. Then, in the assumption that the normalized stellar profile gives the initial distribution of globular clusters, we find that the number of ``missing clusters'' is 380. Hence NGC 1379 could have lost about 74\% of its initial population of globular clusters.\subsubsection{Data completeness and the mass fallen to the galactic centre}The relevant number of  clusters possibly missing in the inner region of NGC 1379  corresponds to a great amount of mass supposedly fallen close to the galactic centre. Remembering that the value obtained of $N_{l}$ is restricted to clusters with $B<25.5,$ we decided to provide two different evaluations of the mass removed from the GCS: one is the missing mass in the magnitude range of the observations, $M_{l},$ and another is the missing mass in the ``whole'' magnitude range, $\tilde{M}_{l}.$ For this purpose it is useful to make some considerations on the GCS luminosity function. 

F98b reported $B$ and $I$ photometry for $\sim 300$ globular clusters candidates in NGC 1379. This sample was detected with 4 WFPC2 chips covering a total area of 4.8 arcmin $^{2}.$ The sample with $B\leq 26$ is almost complete; at $B>26$ the completeness begins to decrease very sharply. . The principal source of contamination by foreground stars and background galaxies is due to compact, spherical galaxies and F98b provided a background-corrected luminosity function as a gaussian function with a peak magnitude of $\langle B\rangle =24.95\pm 0.30$ and a width $\sigma_{B}=1.55\pm 0.21.$ This luminosity function is almost complete and uncontaminated in the range $21<B<25.5$. Since we want to estimate the mass lost by the GCS, we had to convert the magnitude range $21<B<25.5$ in a mass range. 
This was done assuming for the galaxy a distance modulus $m-M=31.32$ for NGC 1379 (Madore, 1986) and , for the GCS, an average colour $\langle B-V \rangle =0.7$ and a mass to light ratio $\left( M/L \right)_{V,\odot }=1.5$ , thus obtaining
($5.2 \times 10^{4}M_{\odot }<m <3.27\times 10^{6}M_{\odot }$). 
Adopting the mass spectrum (3), the mean mass value of the clusters 
(over the told mass rangei) is $\langle m \rangle = 3.5\times 10^{5}M_{\odot },$.
  This value, together with the number of the missing globular clusters (380), leads to $M_{l} = 1.3\times 10^{8}M_{\odot }$ as estimate of the mass fallen to the galactic centre. 

\noindent A discussion of the error in $M_l$ induced by the error  $\frac{\Delta \langle m\rangle }{\langle m\rangle }$ for this agalaxy, as well for NGC 1399 and NGC 1404, is postponed to   Sect. 3. With regard to the value of the mass lost, we remind it represents an understimate of the mass removed from the GCS during its evolution because it is limited to the actual magnitude range of the observations. Hence to compute the ``total'' mass lost we set a minimum and a maximum mass, chosen as $m_{\min }=10^{4}M_{\odot }$ and $m_{\max }=10^{7}M_{\odot }.$ Actually, normalizing the complete mass spectrum to reproduce the number (132) of clusters in the observed range (21$<B<25.5$), we obtain $N_{l,tot}$ value of 179. Moreover, the averaged mass value over the whole magnitude range is 2.9$\times 10^{5}M_{\odot }$ thus to give as estimate of the total missing mass ${M}_{l,tot}=1.5\times 10^{8}M_{\odot }, i.e. $15 \% greater than  the value obtained in the magnitude range of the observations.\subsection{NGC 1399}NGC 1399 is a giant elliptical galaxy lying about at the centre of the Fornax cluster. Among the galaxies around the cluster, NGC 1399 is the most populous in globular clusters. Its GCS was deeply studied in the past through ground-based observations. Among these studies we mention that by Kissler-Patig et al. (1997), who deduced a total number of globular clusters of $5940\pm 570,$ value that corresponds to the high specific frequency $S_{N}=11\pm 4.$ 

In F98a is presented the optical imaging of this galaxy obtained with the HST WFPC2 and the Cerro Tololo Inter-American Observatory 1.5m telescope. Combining these data they provided the spatial distribution of globular clusters used in the present study.\subsubsection{The data and their fitting formulas}
The imaging from HST led F98a to set as completeness limit $B=26.5.$ In order to avoid misleading revelations, they excluded also all the sources brighter than $B=21,$ since they would be more luminous than known normal globular clusters and so, probably, foreground stars. Moreover, these authors introduced a colour cut to the data: 

only the sources with $1.2<B-I<2.5$ have been incuded in their sample. We see that 95\% of all sources lies inside this colour selection. F98a drawn a surface density profile for their sample of globular clusters. To do this, they introduced a small correction ($\sim 1.5$ \%) to take into account  the number of undetected globular clusters, that are expected to be at the faint end of the luminosity function ($B\simeq 26.5)$. Next, they subtracted off the estimated background density of 2.9 objects per arcmin$^{2}.$ Excluding the innermost density points, they fitted the data with the function: \begin{equation}\Sigma \left({ r }\right) =126r^{-1.2\pm 0.2}  \end{equation}\noindent where $r$ and $\Sigma (r)$ are  expressed in arcmin and arcmin$^{-2}$, respectively. It is clear from Fig.2  that the  GCS surface density flattens out in the inner region.  The
estimate of the core radisu given in F98a is 40 arcsec ($3.4$ kpc). We note that this value of the core radius lies in the scatter of the relation found by Forbes et al. (1996) between the GCS core radius and the galaxy luminosity. 

Assuming $40$ arcsec as  core radius, our modified--core model best fit for NGC 1399 GCS is that  with  $\Sigma _{0}=234.42$ arcmin$^{-2}$ and $\gamma =0.61$ (solid line in Fig.2).
As done first by Hanes \& Harris (1986) and later by F98a, we assumed 10 arcmin as limiting radius of the GCS.
 However it must be taken into account that an error on the limiting radius leads to an error in the evaluation of the globular cluster number. F98a estimated the error in the limiting radius $\pm 1$ arcmin; consequently, integrating our fit within 10 arcmin, we find that the total number of globular clusters in NGC 1399 is $6100\pm 540$. 
This value agrees well with the values found in previous studies (Kissler-Patig et al. 
1997 gave $5940\pm 570$ and F98a obtained the value $5700\pm 500,$ adopting the profile (6) completed by  the inner available data).
\begin{figure}
\vspace{1pt}
\hspace{10pt}
\epsfxsize=240pt
\epsfbox{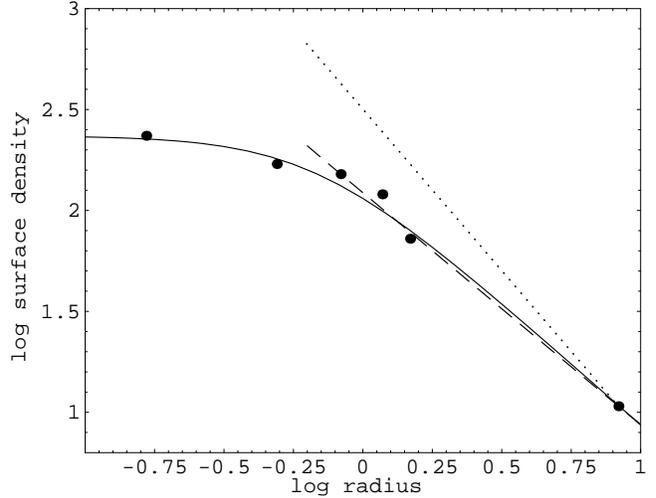}
%\epsfbox{n1399.eps}
\par\caption{Surface density profile for the GCS of NGC 1399 (black dots) from HST data in F98a. Our  modified core model fit is given by the solid line, while  the dashed line represents the power law fit to data given in F98a. The underlying stellar light distribution  arbitrarly normalized in the vertical direction to match the globular cluster system in the outer region is shown as dotted line. Radial distance is in arcsec (1 arcsec = 85 pc); surface density in arcmin$^{-2}$. }
\end{figure}
\subsubsection{Number of missing globular clusters in NGC 1399}The presumed flattening of the GCs distribution in NGC 1399 has been deeply discussed. While Bridges et al. (1991) stated that the GCS has the same slope of the halo stars, Wagner et al. (1991) and Kissler-Patig (1997), on the contrary, noticed a flattening of the GCS radial profile toward the galactic centre. A weighted average of the slopes provided by studies earlier than that of F98a brings to a value of $-1.47\pm 0.09$ for the GCS and $-1.72\pm 0.06$ for the halo stars.\noindent F98a found a difference of $0.4\pm 0.22$ in the slopes between their fitted GCS distribution profile and that of the stellar halo, concluding that the flattening of the GCS density profile is a real feature. 
To provide an estimate of `missing' GCs in NGC 1399, we adopted for the stellar light the following profile
\[\log I(r)=-1.6\log r+2.502, \]
\noindent 
obtained shifting the stellar halo density data (Forbes 1999, private communication) in the vertical direction, such to match the GCS profile in the outer regions. 
Extrapolating inward this profile to the galactic radius of 0.1 arcmin and integratig this surface density distribution in the galactic region where the GCS and the halo stars profiles differ, i.e. $\sim 0.1-8.3$ arcmin, we found a 
number of stars of about 9680.
On the other side, in this radial range we find a number of GCs equal to 5165, which yields to an estimate of 4515 clusters lost, i.e. about 47\% of the initial GC population.

\subsubsection{Data completeness and evaluation of the mass lost}
The previous results are limited to the magnitude range $21<B\;<26.5$ and to the colour range $1.2<B-I<2.5$. We now extend them to fainter magnitudes, obtaining two different evaluations of the mass lost in form of GCs. To this scope, it is necessary to make some considerations about the colour indices of the GCs in NGC\ 1399. Those in the outer region of the galaxy are bluer than those in the central region: $B-I=1.70$ and 1.97, respectively (F98a). F98a made an estimate of the influence of the contamination of 
background galaxies to the globular cluster sample. In this way, they found 1.84 as an average value of $B-I$. This allows us to find a value for $\langle B-V\rangle $, which is needed to transform the magnitude interval in a mass range. We have adopted the relation $\left( B-V\right) _{0}=0.347(B-I)_{0}+0.163$  (Grillmair et al. 1999), obtaining 
$\langle B-V\rangle \simeq 0.8$. 

Adopting a distance modulus $m-M=31.2$ 
(Jacoby et al. 1992) and a colour index $\langle B-V\rangle \simeq 0.8$ for NGC 1399, and a mass-luminosity ratio, for its GCs, equal to 1.5, we obtain that the interval $21<B<26.5$ transforms into the mass range
 $2\times10^{4}M_{\odot }<m<3.2\times 10^{6}M_{\odot }$. The mean value of the GC mass is then $\langle m\rangle =2.86\times 10^{5}M_{\odot }$. 
Since the number of lost globulars is 4515, the mass loss is $1.3\times 10^{9}M_{\odot }$. 

To estimate the mass lost in the whole GC magnitude range, we adopted as total range of globular cluster masses that between $m_{\min}=10^{4}M_{\odot }$ and $m_{\max }=10^{7}M_{\odot }$. 
Normalizing the mass distribution (3) to the number of globular clusters observed in the range $21<B<26.5$, we obtain a total number of 5686 clusters, to compare with the 5165 observed in the magnitude range covered by the observations. This leads to an estimate of the total mass lost of $1.44\times 10^{9}M_{\odot }$, which differs of about 10\% from that previously obtained.
\subsection{NGC 1404}
NGC 1404 is an E1 galaxy, about a magnitude fainter than NGC 1399, sited at about 10 arcmin south east of NGC 1399. 

As for NGC 1399, much work has been done on the GCS of this galaxy. The estimates of the number of globular clusters of NGC 1404 are made difficult by its vicinity to NGC 1399. Grillmair et al. (1997)  speculated that a great number of the NGC 1399 GCs (those of intermediate metallicity) have been tidally stripped by NGC 1404. The difficulties in distinguishing the 
two GCSs did not allow to obtain precise results from earlier studies, 
based on photographic plates (e.g., Hanes \& Harris 1986 and Richtler et al. 1992).
 This problem seems to have been solved by the recent studies by F98a, based on data obtained by the HST WFPC2 and the 1.5 m CTIO telescope.

\subsubsection{The data and their fitting formulas}
As for NGC 1399, the magnitude and colour intervals of the GC sample are $21<B<26.5$ and $1.2<B-I<2.5$.  About 90\% of the observed clusters fall in these ranges. Since the number of background objects in the field is 14, F98a concluded that the completeness of the sample of clusters of NGC 1404 is $\geq 93$\%. The GCS surface density profile of NGC 1404 has been obtained by subtracting both the background density and the density of those clusters which are expected to belong to NGC\ 1399. Excluding the innermost data point, F98a fitted the resulting data with the following surface density profile:
\[\Sigma \left( r\right) =36r^{-1.3\pm 0.2} \]
\noindent
where $r$ and $\Sigma\left(r\right)$ are in arcmin and arcmin$^{-2}$, respectively. Fig.3 shows a central flattening of the GCS distribution, which allows F98a to define, also in this case, a `core' for the system.
 Their estimate of the  core radius is of about 30 arcsec (2.5 kpc). 
As for NGC 1399, this core radius lies in the scatter of the correlation found by Forbes et al. (1996). We fitted the GCS density distribution data with a modified core model, as we have done for the two other galaxies. 
The best fit is that with $\Sigma_{0}=164.39$ arcmin$^{-2},$ $\gamma =0.85$ and $r_{c}=30$ arcsec (solid line in Fig.3).
\noindent To obtain an estimate of the total number of globular clusters in this galaxy,
 we have integrated our density law in the galactic region within 250 arcsec 
(about 4 arcmin), which is the limiting radius of the observed data. 
F98a, in fact, showed that in the direction of NGC 1399, the GCS surface density is clearly dominated by this galaxy beyond 250 arcsec. 

The modified core model gives a number of globular clusters of 744 $\pm 120$ in this galactic region (the error in the GC number has been calculated adopting an error in the limiting radius equal to $\pm 1$ arcmin, as suggested by F98a). Our estimate of the GC total number is close to that of F98a; indeed, using the density profile (8) together with inner data points they obtained a number of GCs equal to 725$\pm 142$. Our value
is between that estimated by Harris \& Hanes (1986), 190$\pm 80,$ and that by 
Richtler et al. (1992), 880$\pm120.$ 

\begin{figure}
\vspace{1pt}
\hspace{10pt}
\epsfxsize=240pt
\epsfbox{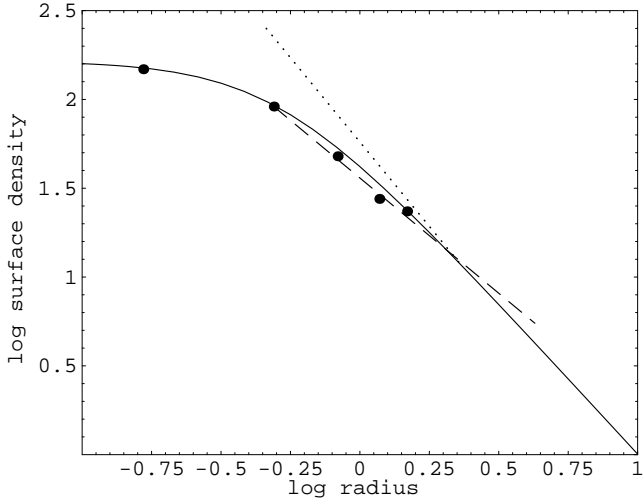}
%\epsfbox{n1404.eps}
\par
\caption{Surface density profile for GCSs in NGC 1404 (black dots) from HST 
data in F98a. The modified core model fit adopted in our study is represented by the solid line, while the power law fit to the same data by F98a is shown as the dashed line. The underlying stellar light (dotted line) arbitrarly normalized in the vertical direction to match the globular cluster system is also shown. Radial distance is in arcsec (1 arcsec = 85 pc); surface density in arcmin$^{-2}$.}
\end{figure}

\subsubsection{Number of missing clusters in NGC 1404}
Also in this case, to compute the number of globular clusters lost to the galactic centre of NGC 1404 it is necessary to discuss the stellar halo profile. 
F98a determined a logarithmic slope of the star profile of -1.9 $\pm 0.1$. 
It is clear that the NGC 1404 GCS distribution is flatter than that of the halo. 
Thanks to data provided to us in private communication by Forbes (1999) we adopted the following halo star profile
\begin{equation}
\log I(r)=-1.9\log r+1.76 
\end{equation}
where $r$ is in arcmin and $I(r)$ in arcmin$^{-2}$, 
which, as usual, was obtained by a vertical normalization so to match the GCS profile in the outer regions (see dotted line in Fig.3). 
Extrapolating this profile inward to the galactic radius of 0.1 arcmin and integrating this surface density distribution in the region of difference with the GCS profile ($\sim 0.1-2.32$ arcmin), we get 1061 as ``initial'' GC number. 
Integrating our GCS density data fit in the same annulus, we find 508 presently observed clusters, then we conclude that 553 GCs disappeared, i.e. $\sim 52\%$ of its initial clusters population.

\subsubsection{Data completeness and evaluation of the number and mass of lost globular clusters}
The data analysed are limited to the magnitude range 21$<B<26.5,$ transformed into a mass range in order to estimate the mass lost from the GCS of NGC 1404. For this aim let us do some considerations on the colour indices of the GCS. According to the analysis of F98a, the HST data show that NGC 1404 is dominated by red globular clusters, with a significant tail to the blue. Their colour distribution yields a mean value $\langle B-I\rangle $ =1.96, but F98a considered necessary to account for the background contamination. Using the CTIO data they found a bluer colour distribution,
 with a mean value $\langle B-I \rangle =1.71$. Now, thanks to the relation between $B-V$ and $B-I$ given by Grillmair et al. (1999) we transformed 
$\langle B-I\rangle $ into $\langle B-V\rangle ,$ obtaining a $\langle B-V\rangle $
 value of about 0.75. This value for $\langle B-V\rangle $ together with 
a distance modulus $m-M=31.2$ maps the magnitude range 21$<B<26.5$ onto the mass range $1.95\times 10^{4}M_{\odot }<m<3\times 10^{6}M_{\odot }$. 
This allows to estimate a GC mean mass $\langle m\rangle = 
2.48\times10^{5}M_{\odot }$ and, being the number of missing clusters equal to 553, we deduce an amount of the mass lost by GCS of 1.37$\times 10^{8}M_{\odot }.$ 

With regard to the total mass lost we have again adopted the mass range 
$10^{4}M_{\odot }<$ $m<10^{7}M_{\odot }$ as total mass range of globular clusters. 
Again, normalizing the mass spectrum (3) to reproduce the number of globular
clusters actually observed in the  magnitude range $21<B<26.5,$ we find 556 as ``total'' number of GCs. 
This number, together with our estimate of the mean mass value in this range 
(2.9$\times 10^{5}M_{\odot })$, gives a total mass lost of 1.75$\times 10^{8}M_{\odot },$ value 27\% greater than the previous one.
\begin{table}
\caption{The presently observed number of  clusters  ($N$), its initial value ($N_i$), its fractional variation ($\Delta$), the mass lost in the magnitude range of the observations ($M_l$) and in the whole magnitude range ($M_{l,tot}$)}.
\begin{tabular}{llllll}
\hline Galaxy & $N$ & $N_{i}$ & $\Delta $ & $M_{l}(M_{\odot })$ & ${M}_{l,tot}(M_{\odot })$ \\
\hline\hline\textbf{NGC 1379} & 132 & 512 & 0.74 & $1.30\times 10^{8}$ & $1.50\times 10^{8}$\\ 
\textbf{NGC 1399} & 5165 & 9680 & 0.47 & $1.30\times 10^{9}$ & $1.44\times10^{9}$ \\ \textbf{NGC 1404} & 508 & 1061 & 0.52 & $1.37\times 10^{8}$ & $1.75\times10^{8}$ \\ 
\hline\end{tabular}\end{table}
\section{The error on the mean mass value}
Using the expressions deduced in Appendix A we can give the error in the estimate of the average value of the mass, $\Delta \langle m \rangle\over \langle m \rangle$, of the GCs in the 3 galaxies studied.

We get
\begin{equation}\frac{\Delta \langle m\rangle }{\langle m\rangle }=-0.05
\frac{\Delta s_{0}}{s _{0}}-1.48\frac{\Delta s _{1}}{s _{1}}-0.21
\frac{\Delta s _{2}}{s _{2}}+0.45\frac{\Delta \beta }{\beta }, 
\end{equation}
\begin{equation}\frac{\Delta \langle m\rangle }{\langle m\rangle }=
-0.15\frac{\Delta s_{0}}{s _{0}}-1.63\frac{\Delta s _{1}}{s _{1}}-0.19
\frac{\Delta s _{2}}{s _{2}}+0.37\frac{\Delta \beta }{\beta }, %\label{7}
\end{equation}
\begin{equation}\frac{\Delta \langle m\rangle }{\langle m\rangle }=
-0.58\frac{\Delta s_{0}}{s _{0}}-1.95\frac{\Delta s _{1}}{s _{1}}-0.19
\frac{\Delta s _{2}}{s _{2}}+0.36\frac{\Delta \beta }{\beta },  
\end{equation}
for NGC 1379, NGC 1399 and NGC 1404, respectively.
From these expressions it is clear that the greatest source of indetermination is due to $s _{1}$, i.e. to the parameter of the mass spectrum that characterizes the intermediate mass range. However it is important to underline that the abundance of the GCs sample in these ranges allows an accurate determination of $s _{1}$. 
Assuming an error of 10\% in the  parameters $s _{0},$ $s _{2}$
 and $\beta \equiv \left( M/L\right) _{V,\odot }$, we obtain, as 
resulting errors in $\langle m\rangle$ 22\%, 23\% and 30\%, for NGC 1379, NGC 1399 and NGC 1404, respectively.
\section{The correlation between the mass lost by GCS and the central galactic black hole mass}
Even if theoretical arguments and observational data are clearly compatible with the explanation of the difference between the GCS and stellar bulge radial distributions as a result of dynamical evolution, we still cannot firmly state it.  
By the way it is interesting, and surely constitutes an argument in favour of this interpretation, to see how our results for the (supposed) mass loss in form of glbobular cluster correlates with the galactic central black hole mass, at least for the set of seven galaxies for which these data are available (our galaxy, M 31, M 87, NGC 1399, NGC 4365, NGC 4589, IC 1459). 

\noindent Sources of data in Fig. 4 are: for $M_l$ this paper and Capuzzo--Dolcetta \& Vignola, 1997, while the black hole masses are from Magorrian et al. (1998) except for our Galaxy and IC 1459 taken from Gebhardt et al. (2000). 
A recent estimate by Saglia et al. (2000) gives as compatible with kinematic data of the inner 5 arcsec of  NGC 1399 a black hole mass $\sim 5\times 10^8 M_\odot$, i.e. about a  factor of 10 less than the value estimated by Magorrian et al. (1998) and reported in Fig. 4. 

 Fig. 4 shows, even in the paucity of data available, a clear increasing trend of $M_l$ as function of $M_{bh}$. Actually, the least square straight line fitting the data relative to galaxies with black hole masses heavier than $10^8$ M$_\odot$ has slope $0.99$, i.e. the $M_l$ -
$M_{bh}$ relation is remarkably linear, while the straight line connecting the 2 low mass points has slope $0.08$. The best simple fit to the whole set of data is given by $Log M_l=a+bexp(Log M_{bh})$ with $a=7.158$ and $b=1.194 \times 10^{-4}$, giving $r^2=0.76$ and standard error $=0.44$ (the least square straight line fitting the whole data set yields $r^2=0.61$ and standard error $=0.56$).
Incidentally, we note that the steeper slope of the $M_l$ - $M_{bh}$ relation for high values of the nucleus mass is what expected when the black hole is very massive and it acts as powerful engine of GCS evolution. 

\begin{figure}
\vspace{1pt}
\hspace{10pt}
\epsfxsize=240pt
\epsfbox{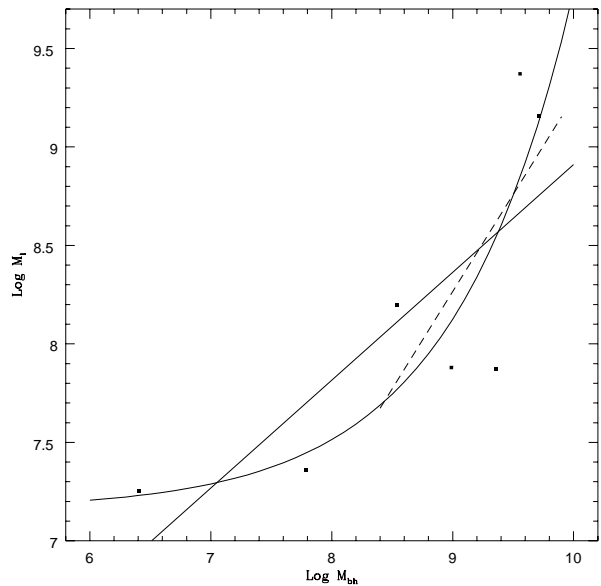}
\par\caption{The correlation between the (logarithmic) GCS mass lost and the central galactic black hole mass for 7 galaxies for which these data are available. Masses are in solar masses.
The  straight solid line is the least square fit to all data,
the dashed line is the least square fit restricted to data of the most massive black holes ($Log M_{bh} > 8$) while the curve is the exponential fit described in the text}.
\end{figure}
\section{Conclusions} 
We used data taken with the WFPC2 of the Hubble Space Telescope to compare the globular cluster system distribution to that of bulge--halo stars in a sample of three elliptical galaxies in the Fornax cluster.

The first relevant outcome of this study is that the radial distributions of the globular cluster systems in NGC 1379, NGC 1399 and NGC 1404 confirm that  the GCSs in elliptical galaxies are less concentrated to the inner regions than the halo stars. Following the scheme of previous works (McLaughlin 1995, Capuzzo-Dolcetta \& Vignola 1997, Capuzzo-Dolcetta \& Tesseri 1999), 
i.e. assuming that the globular clusters and halo star distributions were initially the same and the present difference is due to evolutionary effects acting on the GCS, we computed the number of globular clusters lost as the integral of the difference between the two surface distributions (suitably normalized). 
\par\noindent In this way we have found that the GCSs of NGC 1379, NGC 1399 and NGC 1404 could have lost the 74\%, 47\% and 53\% of their initial populations, respectively. 
These results are, obviously, limited to the magnitude range of the observations within which the data reach a sufficiently high level of completeness. 
\par\noindent  The large number of `missing' globular clusters obtained  reveals that a significant amount of mass could have reached the galactic centre, too. An estimate of the mass fallen to the galactic center needs the assumption of a value for the  initial mean mass of globular clusters, $\langle m\rangle .$ It is reliable (see Capuzzo-Dolcetta \& Tesseri 1997) to assume for $\langle m\rangle $ the mean mass value of the presently observed globular clusters. This quantity can be estimated  assuming a mass spectrum, that we took  as a piece-wise power law for it seems to represent one of the best approximations for elliptical galaxies globular cluster systems (McLaughlin 1994; Harris \& Pudritz 1994). In particular, we adopted as unique mass spectrum for the three galaxies of the Fornax cluster that obtained by normalization of the luminosity function  adopted by McLaughlin (1995) for the Virgo elliptical giant M87, and assuming   a mass to 
light ratio $\left( M/L\right) _{V,\odot }$ =1.5. 
\indent We provided two different estimates of the mass lost in form of globular clusters: the first is that due to clusters in the magnitude range of the observations and the other refers the ``presumed'' whole magnitude range. The first estimate, $M_{l},$ is simply obtained as the product of the mean mass and the number, $N_{l}$, of the missing globular clusters. 
\noindent With regard to the second estimate we made the, quite reasonable, 
assumption that the total number of missing globular clusters scales with the total number of the clusters present in the galaxy. Hence we computed this total number of globular clusters assuming  
$10^{4}M_{\odot }<m<10^{7}M_{\odot }$ as whole mass range, so to include the contribution to $N_{l}$ coming from  unobserved clusters. The lower limit of the mass range has not a big influence on the total mass; with regard to the upper limit, the large decreasing slope ($-3$) of the MF at large masses ensures a high level of convergence of the total mass reached when assuming $10^{7}M_{\odot }$ as maximum mass.   The values of the total mass lost, $\tilde{M}_{l},$ obtained in this way for the three galaxies NGC 1379, NGC 1399 and NGC 1404 are not too different  from those estimated in the mass ranges of the observations (the difference is about 15 \%, 10 \% and 27 \%, respectively; see Table 1). 
This is explained by that the range of magnitude covered by the observations is close enough to the whole magnitude range, thanks to the high power of HST. 
\par It is relevant noting the positive correlation we found, for the available set of 7 galaxies in Fig. 4, between our evaluated mass lost from the GCS system and the central black hole mass of the parent galaxy: this is what expected on the basis of globular cluster-black hole tidal interaction because the more massive the central black hole the greater its influence on the GCS population.
\par To conclude, we remark how the values of the mass lost in form of decayed and destroyed globular clusters is significant and so it should have been played a role in the central region activity of their parent galaxies.

\section*{Acknowledgments}
We warmly thank D.A. Forbes for giving us some unpublished data useful for the determination of the stellar profile in NGC 1404.
We are grateful to D.A. Forbes and C.J. Grillmair  for important discussions and comments about the data used in this work.

\appendix\section{}
A significant source of error on the evaluation of the mass removed from the GCS is due to the assumption of the exponent $s$ of the mass spectrum and of the value of the mass-light ratio $\left( M/L\right) _{\odot }$ (hereafter $\beta $). Infact, an error on these parameters leads to the relative error on $\langle m\rangle $
\begin{equation} \frac{\Delta \langle m\rangle }{\langle m\rangle }=\frac{\partial \langle m\rangle }{\partial s }\frac{s }{\langle m\rangle }\frac{\Delta s }{s }+\frac{\partial \langle m\rangle }{\partial \beta }\frac{\beta }{\langle m\rangle }\frac{\Delta \beta }{\beta }  
\end{equation} 
Before giving an explicit expression of (A1), we  note that, if we use the mass spectrum as in (3), $\langle m\rangle $ is so defined:
\begin{equation}
\langle m\rangle =\langle \beta L\rangle =\beta \langle L\rangle =\beta \frac{\int_{L_{\min }}^{L_{\max }}k(L)L^{-s +1}dL}{\int_{L_{\min}}^{L_{\max }}k(L)L^{-s }dL}  
\end{equation}
that is, in terms of mass
\begin{equation}\langle m\rangle =\frac{\int_{m_{\min }}^{m_{\max }}k(m)m^{-s +1}dm}{\int_{m_{\min }}^{m_{\max }}k(m)m^{-s }dm}  
\end{equation} 
where $k(m)$ is obtained by the proper tranformation of the McLaughlin (1994) normalizing factor $k(L)$, i.e.
\begin{equation}k(m)=\left\{ \begin{tabular}{ll}$A$ & $m_{\min }\leq m\leq m_{1}$ \\ $Am_{1}^{s _{1}-s _{0}}$ & $m_{1}\leq m\leq m_{2}$ \\ $Am_{1}^{s _{1}-s _{0}}m_{2}^{s _{2}-s _{1}}$ & $m_{2}\leq m\leq m_{\max }$\end{tabular}.\right.  
\end{equation}
where, of course, $m_{min}$, $m_{max}$, $m_{1}$ and $m_{2}$ depend linearly on $\beta$. To give (A1) a more compact form, we have put 
\begin{description}\item  $m_{0}\equiv m_{\min }$ and $m_{3}\equiv m_{\max },$\item  $k_{0}\equiv A,$ $k_{1}\equiv Am_{1}^{s _{1}-s _{0}},$ $k_{2}\equiv Am_{1}^{s _{1}-s _{0}}m_{2}^{s _{2}-s _{1}},$\item  $N_{tot}=\int_{m_{\min }}^{m_{\max }}k(m)m^{-s }dm.$
\end{description}
\noindent Remembering that $s =\left( s _{0},s _{1},s_{2}\right) $ (A1) becomes
\begin{equation}\frac{\Delta \langle m\rangle }{\langle m\rangle }=\sum_{i=0}^{2}\frac{\partial \langle m\rangle }{\partial s _{i}}\frac{s _{i}}{\langle m\rangle }\frac{\Delta s _{i}}{s _{i}}+\frac{\partial \langle m\rangle }{\partial \beta }\frac{\beta }{\langle m\rangle }\frac{\Delta\beta }{\beta }  
\end{equation}
where $\frac{\partial \langle m\rangle }{\partial s _{i}},$ $\frac{\partial \langle m\rangle }{\partial \beta }$ are so defined:
\begin{equation}\frac{\partial \langle m\rangle }{\partial s _{i}}{\footnotesize =}\frac{1}{N_{tot}}\sum_{j=0}^{2}\left[ 
\begin{array}{c}\left( \varphi _{i}^{j}\left( j+1,2\right) -\varphi _{i}^{j}\left(j,2\right) \right) + \\ -\langle m\rangle \left( \varphi _{i}^{j}\left( j+1,1\right) -\varphi_{i}^{j}\left( j,1\right) \right)\end{array}\right]  \end{equation} with $\varphi _{i}^{j}(r,s)=\frac{\partial k_{j}}{\partial s _{i}}\frac{m_{r}^{-s _{j}+s}}{-s _{j}+s}+\delta _{i}^{j}\frac{m_{r}^{-s_{j}+s}}{-s _{j}+s}k_{j}\left( \frac{1}{-s _{j}+s}-\ln m_{r}\right) ,$and
\begin{equation}\frac{\partial \langle m\rangle }{\partial \beta }=\frac{1}{N_{tot}}\sum_{j=0}^{2}\frac{k_{j}}{\beta }\left[ 
\begin{array}{c}\left( m_{j+1}^{-s ^{j}+2}-m_{j}^{-s ^{j}+2}\right) + \\ -\langle m\rangle \left( m_{j+1}^{-s ^{j}+1}-m_{j}^{-s^{j}+1}\right)\end{array}\right] .  \end{equation}
\end{document}